\documentclass[preprintnumbers,showpacs,amsmath,amssymb,floatfix,9pt,prd,onecolumn,
superscriptaddress,nofootinbib]{revtex4}
\usepackage{latexsym}
\usepackage{epsfig}
\usepackage{amssymb}

\begin{document}

\title{Exact solutions in $(2+1)$-dimensional anti-de Sitter space-time admitting a linear or non-linear equation of state}

\author{Ayan Banerjee }
\email{ayan_7575@yahoo.co.in } \affiliation{Department of
Mathematics, Jadavpur University, Kolkata 700032, West Bengal,
India.}
\author{Farook Rahaman}
\email{rahaman@iucaa.ernet.in} \affiliation{Department of
Mathematics, Jadavpur University, Kolkata 700032, West Bengal,
India.}
\author{Kanti Jotania}
\email{kanti@iucaa.ernet.in}
\affiliation {Department of Physics, Faculty of Science, The M. S.
University of Baroda, Vadodara 390002, India.}
\author{Ranjan Sharma}
 \email{rsharma@iucaa.ernet.in}
\affiliation {Department of Physics, P. D. Women's College,
Jalpaiguri 735101, India.}
\author{Mosiur Rahaman}
\email{mosiurju@gmail.com}
\affiliation {Department of Mathematics, Meghnad Saha Institute of
Technology, Kolkata 700150, India.}

\begin{abstract}
Gravitational analyzes in lower dimensions has become a field of
active research interest ever since Ba$\tilde{n}$ados, Teitelboim
and Zanelli (BTZ) ({\em Phys. Rev. Lett.} {\bf69}, 1849, 1992)
proved the existence of a black hole solution in $(2+1)$ dimensions.
The BTZ metric has inspired many investigators to develop and
analyze circularly symmetric stellar models which can be matched to
the exterior BTZ metric. We have obtained two new classes of
solutions for a $(2+1)$-dimensional anisotropic star in anti-de
Sitter  background space-time which have been obtained by assuming
that the equation of state (EOS) describing the material composition
of the star could either be linear or non-linear in nature. By
matching the interior solution to the BTZ exterior metric with zero
spin, we have demonstrated that the solutions provided here are
regular and well-behaved at the stellar interior.
\end{abstract}


\maketitle

\section{Introduction}

Lower dimensional gravity, due to its comparatively simpler setting,
plays a crucial role towards our understanding of many conceptual
issues relating to Einstein's gravity. Gravitational analyzes in
three dimensions got a tremendous impetus when Ba\~{n}ados,
Teitelboim and Zanelli \cite{BTZ}(henceforth BTZ) proposed a model
for a circularly symmetric charged body in an anti-de Sitter back
ground space-time which was found to admit a black hole solution in
the presence of a negative cosmological constant. The BTZ metric is
characterized by its mass, angular momentum and charge.

The existence of a black hole in $(2+1)$ dimensions has inspired
many investigators to construct and study circularly symmetric
star models. Different techniques have so far been adopted
to generate static interiors solutions corresponding to the BTZ
exterior metric. For example, Cruz and Zanelli \cite{Cruz} have
obtained an exact solution for an incompressible fluid in $(2+1)$
dimensions. For a given density profile, Cruz {\em et al}
 \cite{Cruz2} have obtained new class of solutions corresponding to
exterior BTZ metric. Cataldo and Salgado \cite{Cataldo} have
analyzed an Einstein-Maxwell system in $(2+1)$ dimensions. For a
polytropic equation of state (EOS),
 Paulo M. S$\acute{a}$ \cite{Paulo} has proposed a formalism to obtain interior solutions corresponding to the BTZ exterior metric. Sharma
  {\em et al} \cite{Sharma} have assumed a particular mass function to obtain new class of solutions in $(2+1)$ dimensions. Garc\'{i}ýa and
 Campuzano \cite{Garc} have proposed a formalism to obtain circularly symmetric solutions from known density profile or EOS of the fluid source.
 Making use of Finch and Skea \cite{Finch} ansatz in $(2+1)$ dimensions, Banerjee {\em et al} \cite{Ayan} have generated new class of physically
  acceptable interior solutions corresponding to the BTZ exterior. Rahaman {\em et al} \cite{Farook} have studied the properties of BTZ black hole
   by proposing new exact solutions of Einstein's field equations in $(2+1)$ dimensional anti-de Sitter back ground space-time in the context of
    non-commutative geometry.

Motivated by such developments, we propose here two new classes of exact solutions describing the interior of a circularly symmetric
 star with zero angular momentum in an anti-de Sitter back ground space-time. In our construction, we assume that the material composition
 of the star is anisotropic in nature and generate solutions for linear as well as non-linear EOS. We analyze physical behaviour of the model
  by matching the interior solution to the BTZ exterior metric with zero spin and show that the solutions generated here are regular and well-behaved
   at the stellar interior.

\section{Interior space-time}
We write the line element for a static circularly symmetric star with zero angular momentum in the form
\begin{equation}
ds^2 = -e^{2\nu(r)}dt^{2}+ e^{2\mu(r)}dr^{2} + r^{2} d\theta^{2},\label{eq1}
\end{equation}
where $\nu(r)$ and $\mu(r)$ are yet to be determined. The Einstein's field equations for an anisotropic fluid in the
 presence of a negative cosmological constant $(\Lambda < 0)$ are then obtained as (we set $G = c = 1$)
\begin{eqnarray}
2\pi\rho +\Lambda &=& \frac{\mu^{\prime}e^{-2\mu(r)}}{r},\label{eq2}\\
2\pi p_r -\Lambda &=& \frac{\nu^{\prime}e^{-2\mu(r)}}{r},\label{eq3}\\
2\pi p_t -\Lambda &=& e^{-2\mu}\left({\nu^{\prime\prime}}+{\nu^{\prime 2}}-\nu^{\prime}\mu^{\prime}\right),\label{eq4}
\end{eqnarray}
where, $\rho$ is the energy density, $p_r$ is the radial pressure and $p_t$ is the tangential pressure. Eqs.~(\ref{eq2})-(\ref{eq4}) may be combined to yield
\begin{equation}
\left(\rho+p_r\right)+{p_r^{\prime }}+\frac{1}{r} \left(p_r-p_t\right) = 0,\label{eq5}
\end{equation}
which is analogous to the generalized Tolman-Oppenheimer-Volkoff (TOV) equation in $(3+1)$ dimensions. Defining the mass within a radius $r$ as
\begin{equation}
m(r)= \int^{r}_0 2\pi \rho ~\widetilde{r}~d\widetilde{r},\label{eq6}
\end{equation}
Eq.~(\ref{eq2}) yields
\begin{equation}
2m(r) = C-e^{2\mu(r)}-\Lambda r^2,\label{eq7}
\end{equation}
where $C$ is integrating constant. Following an earlier treatment\cite{Sharma}, we set $C=1$ and assume $2\mu(r)= Ar^2$ so as to
 ensure regular behaviour of the mass function $m(r)$ at the centre. The energy density is then obtained as
\begin{equation}
\rho = \frac{1}{2\pi}\left[Ae^{-Ar^2}-\Lambda\right].\label{eq8}
\end{equation}
The constant $A$ can be determined from the central density
\begin{equation}
\rho_c = \rho(r=0)=\frac{1}{2\pi}\left[A-\Lambda\right].\label{eq9}
\end{equation}
To determine the unknown metric potential $\nu(r)$, we presribe an EOS corresponding to the material composition of the star in the form
\begin{equation}
p_r = p_r\left(\rho,\alpha_1,\alpha_2\right),\label{eq10}
\end{equation}
where $\alpha_1$ and $\alpha_2$ are two positive arbitrary constants constraining the EOS. The physical radius $R$ of the star can be obtained by ensuring that
\begin{equation}
p_r\left(\rho(R),\alpha_1,\alpha_2\right) = 0.\label{eq11}
\end{equation}
The EOS (\ref{eq10}) can be either linear or non-linear in nature and accordingly we consider the two possibilities separately.

\subsection{Case I: Solution admitting a linear EOS}
Let us first assume a linear EOS of the form
\begin{equation}
p_r = \alpha_1\rho+\alpha_2.\label{eq12}
\end{equation}
For the choice (\ref{eq12}), the system (\ref{eq2})-(\ref{eq4}) can be solved analytically and we get
\begin{eqnarray}
\nu(r) &=& \frac{\alpha_1A}{2}r^2-\left(\frac{\alpha_1\Lambda+\Lambda-2\pi\alpha_2}{2A}\right)e^{Ar^2}+C_1,\label{eq13}\\
p_r &=& \frac{\alpha_1}{2\pi}\left[Ae^{-Ar^2}-\Lambda\right]+\alpha_2,\label{eq14}\\
p_t &=& \frac{1}{2\pi} \left[r^2e^{-Ar^2}\left(\alpha_1A-\left(\alpha_1\Lambda+\Lambda-2\pi\alpha_2\right)^2 e^{2Ar^2} +
 \frac{\alpha_1 A}{r^2}-\alpha_1 A^2\right)\right. \nonumber\\
&&\left. -\left(\alpha_1\Lambda+\Lambda-2\pi\alpha_2\right)\left(1+Ar^2\right)+\Lambda \right].\label{eq15}
\end{eqnarray}
where $C_1$ is integrating constant. We also have
\begin{equation}
 \Delta  = p_t-p_r=
\frac{1}{2\pi}\left[r^2e^{-Ar^2}\left(\left(\alpha_1A-(\alpha_1\Lambda+\Lambda-2\pi\alpha_2)e^{Ar^2}\right)^2
-\alpha_1A^2\right)-\left(\alpha_1\Lambda+\Lambda-2\pi\alpha_2\right)Ar^2\right],\label{eq16}
\end{equation}
which is the measure of anisotropy. Note that the anisotropy vanishes at the centre which is a desirable feature of a realistic star.

\subsection{Case II: Solution admitting a non-linear EOS}
Assuming a non-linear EOS of the form
\begin{equation}
p_r = \gamma_1\rho+\frac{\gamma_2}{\rho},\label{eq17}
\end{equation}
where $\gamma_1$ and $\gamma_2$ are positive arbitrary constants, we solve the system (\ref{eq2})-(\ref{eq4}) and obtain
\begin{eqnarray}
\nu(r) &=& \frac{\gamma_1A}{2}r^2-\left(\frac{(\gamma_1+1)\Lambda}{2A}\right)e^{Ar^2}-\frac{2\pi^2\gamma_2}
{A\Lambda}\left(e^{Ar^2}+\frac{A}{\Lambda}\ln(A-\Lambda e^{Ar^2})\right)+C_2,\label{eq18}\\
p_r &=& \frac{\gamma_1}{2\pi}(Ae^{-Ar^2}-\Lambda)+\frac{2\pi\gamma_2}{\left(A e^{-Ar^2}-\Lambda\right)},\label{eq19}\\
p_t(r) &=& \frac{1}{2\pi} \left[\left( \left(\gamma_1-1\right)Ar^2+1\right)\gamma_1 Ae^{-Ar^2}-\Lambda (\gamma_1+1)(1+2Ar^2)
     \right.\nonumber\\
&& \left. +A(1-2\gamma_1)
      \left(\Lambda \gamma_1+\Lambda-\frac{4\pi^2\gamma_2}{Ae^{-Ar^2}-\Lambda}\right)r^2+\left(\Lambda \gamma_1+
      \Lambda-\frac{4\pi^2\gamma_2}{Ae^{-Ar^2}-\Lambda}\right)^2r^2e^{Ar^2} \right. \nonumber\\
&&        \left.
         +4\pi^2\gamma_2\left(\frac{1+2Ar^2}{Ae^{-Ar^2}-\Lambda}+\frac{2A^2r^2e^{-Ar^2}}
         {(Ae^{-Ar^2}-\Lambda)^2}\right)r^2+\Lambda \right].\label{eq20}
\end{eqnarray}
In Eq.~(\ref{eq18}), $C_2$ is an integration constant.

The measure of anisotropy in this case turns out to be
\begin{eqnarray}
 \Delta = p_t-p_r=\frac{1}{2\pi} \left[\left( \gamma_1-1\right)\gamma_1 A^2r^2e^{-Ar^2}-2\Lambda (\gamma_1+1)Ar^2+A(1-2\gamma_1)
      \left(\Lambda \gamma_1+\Lambda-\frac{4\pi^2\gamma_2}{Ae^{-Ar^2}-\Lambda}\right)r^2
     \right.\nonumber\\
        \left. +\left(\Lambda \gamma_1+\Lambda-\frac{4\pi^2\gamma_2}{Ae^{-Ar^2}-\Lambda}\right)^2r^2e^{Ar^2}
         +4\pi^2\gamma_2\left(\frac{2Ar^2}{Ae^{-Ar^2}-\Lambda}+\frac{2A^2r^2e^{-Ar^2}}{(Ae^{-Ar^2}-\Lambda)^2}\right) \right],\label{eq21}
\end{eqnarray}
which vanishes at the centre.

\section{Exterior space-time and boundary conditions}
We assume that the exterior space-time of our circularly symmetric star is described by the BTZ metric (with zero angular momentum)
\begin{equation}
ds^2 = -\left(-M_0 - \Lambda{r^2} \right)dt^2 + \left(-M_0 -
\Lambda{r^2} \right)^{-1}dr^2 + r^2d\theta^2,\label{eq22}
\end{equation}
where $M_0$ is the conserved charge associated with asymptotic
invariance under time displacements. At the boundary  $r=R$,
continuity of the metric potentials yield the following junction
conditions:
\begin{eqnarray}
e^{2\nu(R)} &=& -M_0-\Lambda R^2,\label{eq23}\\
e^{-2\mu(R)} &=& -M_0-\Lambda R^2.\label{eq24}
\end{eqnarray}
Moreover, the radial pressure must vanish at the boundary, i.e., $p_r(r=R) = 0$. These three conditions can be utilized to fix the
values of the  constants $A$, $C$ and $R$ for two different cases, namely solution admitting a linear EOS (Case I) and solution admitting a non-linear EOS (Case II).

\subsection{Case I}
\begin{eqnarray}
A &=& -\frac{1}{R^2}\ln \left(-M_0-\Lambda R^2\right),\label{eq25}\\
C_1 &=& \frac{\alpha_1\Lambda+\Lambda-2\pi\alpha_2}{2A\left(-M_0-\Lambda
R^2\right)}-\left(\frac{1+\alpha_1}{2}\right){AR^2},\label{eq26}\\
R &=& \frac{1}{\sqrt{A}}\left[\ln\left(\frac{A\alpha_1}{\alpha_1\Lambda-2\pi\alpha_2}\right)\right]^{\frac{1}{2}}.\label{eq27}
\end{eqnarray}

\subsection{Case II}
\begin{eqnarray}
A &=& -\frac{1}{R^2}\ln \left(-M_0-\Lambda R^2\right),\label{eq28}\\
C_2 &=& \frac{\Lambda(\gamma_1 +1)}{4A(-M_0-\Lambda R^2)}-\frac{\Lambda
R^2(\gamma_1 +1)}{2}+\frac{2\pi^2\gamma_2}{A\Lambda}\left[\frac{ 1}{
(-M_0-\Lambda R^2)} +\frac{A}{\Lambda}\ln \left( A- \frac{\Lambda
}{ (-M_0-\Lambda R^2)}\right)\right],\label{eq29}\\
R &=& \frac{1}{\sqrt{A}}\left[\ln\left(\frac{A \sqrt{\gamma_1}}{\Lambda
\sqrt{\gamma_1}+\sqrt{-4\pi^2\gamma_2}}\right)\right]^{\frac{1}{2}}.\label{eq30}
\end{eqnarray}

\section{Physical acceptability and regularity of the model}

For a  physically acceptable model, we impose the following restrictions:\\
\begin{itemize}
\item Regularity of the curvature invariants.
\item Energy-density and pressure should be monotonically  decreasing functions of  $r$.
\item Radial sound speed and transverse sound speed should be less than unity i.e., \\ $0 < v_{sr}^2 (=\frac{dp_r}{d\rho}) < 1$,~~
~ $ 0 < v_{st}^2 (=\frac{dp_t}{d\rho}) < 1$.
\end{itemize}

\begin{figure}[htbp]
\centering
\includegraphics[scale=.4]{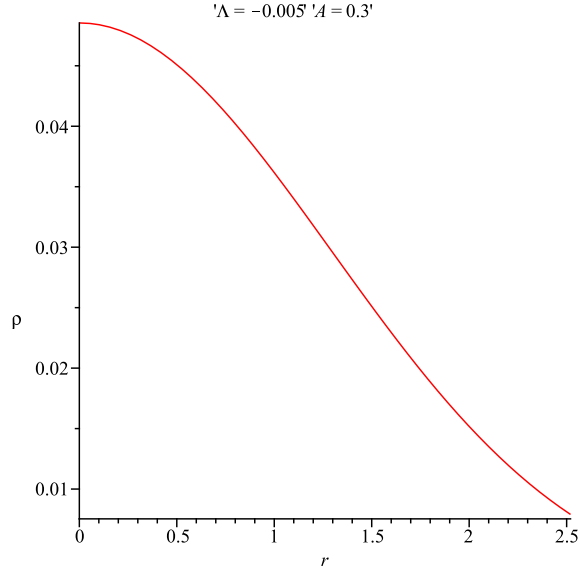}
\caption{Energy density plotted against the radial parameter $r$. }
\label{fig1}
\end{figure}

\begin{figure}[htbp]
\centering
\includegraphics[scale=.4]{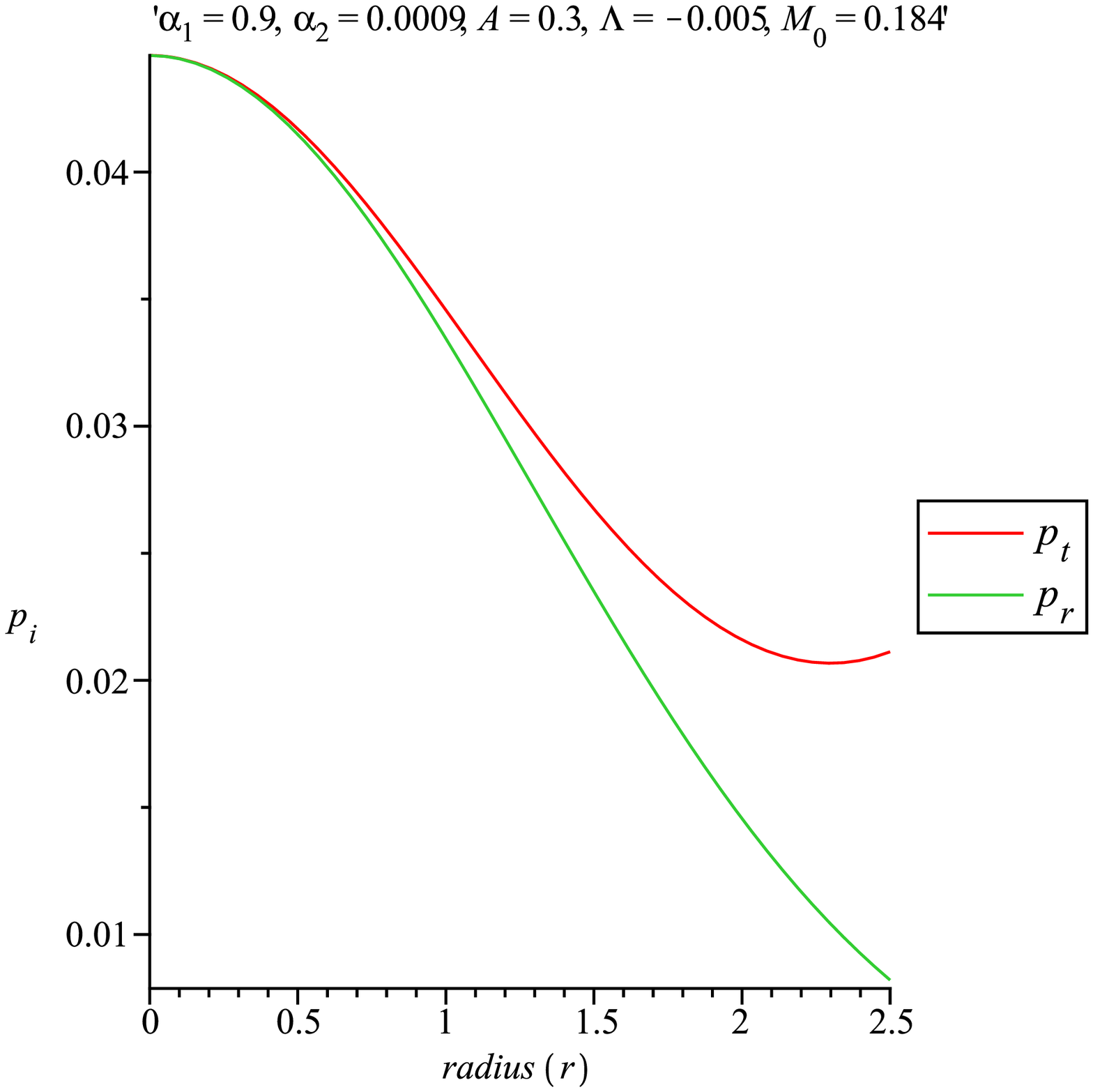}
\caption{Radial and transverse pressures are  plotted against the
radial parameter $r$ for linear EOS.} \label{fig2}
\end{figure}

\begin{figure}[htbp]
\centering
\includegraphics[scale=.4]{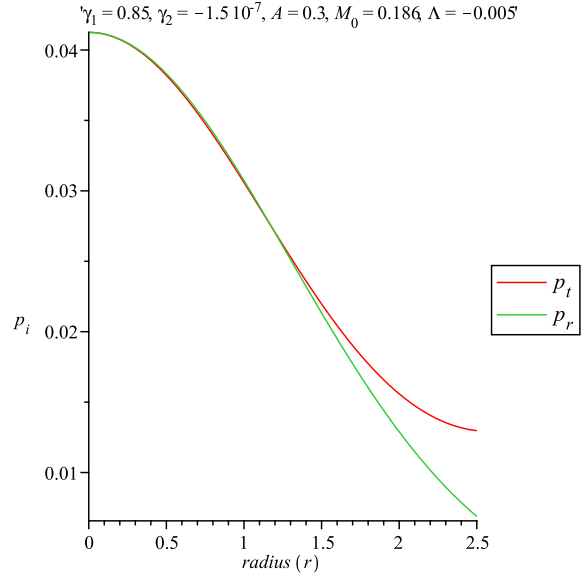}
\caption{Radial and transverse pressures are  plotted against the
radial parameter $r$ for non linear EOS.} \label{fig3}
\end{figure}

\subsection{Case I}

From Eqs.~(\ref{eq5}) and (\ref{eq6}), we obtain
\begin{eqnarray}
\frac{d\rho}{dr} &=& -\frac{A^2}{\pi}re^{-Ar^2},\label{eq31}\\
\frac{dp_r}{dr}&=& -\frac{A^2\alpha_1}{\pi}re^{-Ar^2}, \label{eq32}\\
\frac{d^2 \rho}{dr^2}\Bigl\lvert_{r=0} &=& -\frac{A^2}{\pi}  < 0, \label{eq33}\\
\frac{d^2 p_r}{dr^2} \Bigl\lvert_{r=0} &=& -\frac{A^2\alpha_1}{\pi}< 0. \label{eq34}
\end{eqnarray}
Eqs.~(\ref{eq31})-(\ref{eq34}) show that, for $\alpha_1 > 0$, both
energy-density and radial pressure decrease from their maximum
values at the centre. Variations of energy-density and the two
pressures   have been shown in Fig.~(\ref{fig1})  and
Fig.~(\ref{fig2}), respectively. The Ricci scalar assumed the following form
\begin{equation}
R =2e^{-Ar^2}\left[(1-2\alpha_1)A+(1-\alpha_1)\alpha_1A^2r^2-\left(\alpha_1\Lambda+\Lambda-2\pi\alpha_2\right)^2r^2e^{2Ar^2}+
\left(\alpha_1\Lambda+\Lambda-2\pi\alpha_2\right)\left(2+(1+2\alpha_1)Ar^2\right)e^{Ar^2}\right].\label{eq35}
\end{equation}
One can note that R is regular at the origin and well behaved in the stellar interior.

\subsection{Case II}
Here,
\begin{eqnarray}
\frac{d\rho}{dr} &=& -\frac{A^2}{\pi}re^{-Ar^2},\label{eq36}\\
\frac{dp_r}{dr}&=& -\frac{A^2\gamma_1}{\pi}re^{-Ar^2}+\frac{4\pi A^2\gamma_2}{(Ae^{-Ar^2}-\Lambda)^2}re^{-Ar^2}. \label{eq37}\\
\frac{d^2 \rho}{dr^2}\Bigl\lvert_{r=0} &= &-\frac{A^2}{\pi}  < 0, \label{eq38}\\
\frac{d^2 p_r}{dr^2} \Bigl\lvert_{r=0} &= &
-\frac{A^2\gamma_1}{\pi}+\frac{4\pi A^2\gamma_2}{(A-\Lambda)^2} < 0.
\label{eq39}
\end{eqnarray}
Though it is not straight forward, we note that Eq.~(\ref{eq39})
holds for appropriate choices of the values of $\gamma_1$ and
$\gamma_2$.
 Eqs.~(\ref{eq36})-(\ref{eq39}) show that both energy-density and radial pressure decrease from their maximum values at the centre. Behaviour of
two pressures have been shown in Fig.~(\ref{fig3}) and the Ricci scalar is given by
\begin{eqnarray}
R &=& 2e^{-Ar^2}\left[(1-2\gamma_1)A+\gamma_1A^2r^2+\left((\gamma_1+1)\Lambda+\frac{4\pi^2\gamma_2}{\Lambda}\right)
(1+Ar^2)e^{Ar^2}-\frac{4\pi^2\gamma_2A}{\Lambda\left(A-\Lambda e^{Ar^2}\right)}~~\mathbf{x} \right. \nonumber\\
&&        \left.
\left(1-Ar^2+\frac{A(1+2Ar^2)-\Lambda e^{Ar^2}}{(A-\Lambda e^{-Ar^2})}\right)e^{Ar^2}-
\left( \gamma_1 Ar-\left((\gamma_1+1)\Lambda+\frac{4\pi^2\gamma_2}{\Lambda}\right)re^{Ar^2}
+\frac{4\pi^2\gamma_2 A}{\Lambda(A-\Lambda e^{Ar^2})}re^{Ar^2}\right)^2\right] \label{eq40}
\end{eqnarray}
which is also regular at the stellar interior.
Another important `physical acceptability condition' is the causal property of the radial and tangential
sound speeds which have been addressed in the following sub-sections.

\subsection{Case I}
\begin{figure}[htbp]
\centering
\includegraphics[scale=.4]{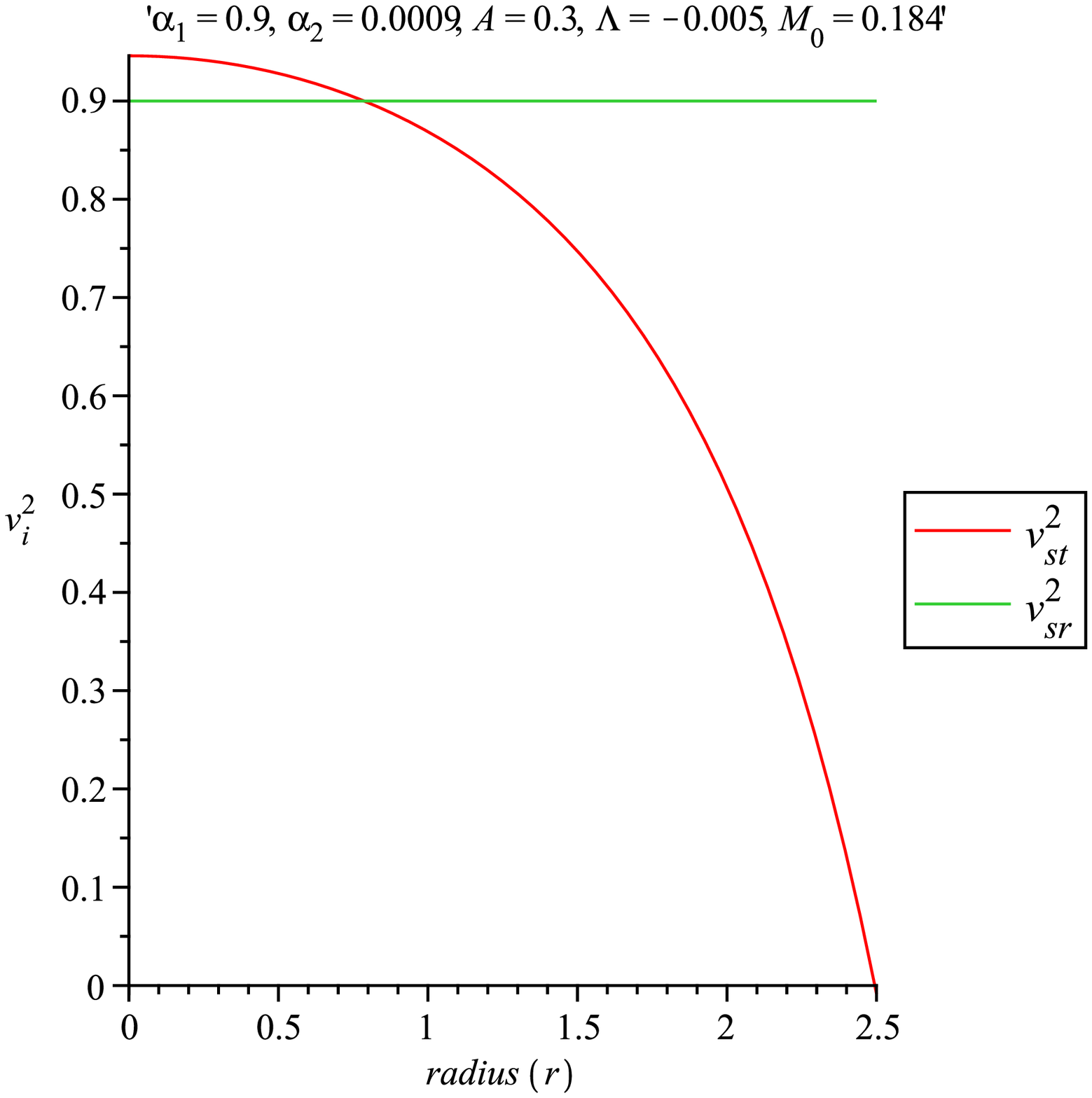}
\caption{Radial and tangential sound speeds plotted against $r$
(Case I).} \label{fig4}
\end{figure}

Combining Eqs.~(\ref{eq31})-(\ref{eq34}), we obtain
\begin{eqnarray}
v^2_{sr} &=& \frac{dp_r}{d\rho} = \alpha_1,\label{eq41}\\
v^2_{st} &=& \frac{dp_t}{d\rho} =
1-(\alpha_1^2-\alpha_1)(1-Ar^2)+(2\alpha_1+1)\left(\frac{\alpha_1\Lambda+\Lambda-2\pi\alpha_2}{A}\right)e^{Ar^2}
  -\left(\frac{\alpha_1\Lambda+\Lambda-2\pi\alpha_2}{A}\right)^2(1+Ar^2).\label{eq42}
\end{eqnarray}
In Fig.~(\ref{fig4}), we have shown the nature of two sound speeds for specific
choices of the model parameters which clearly shows regular behaviour of both the radial and
transverse sound speeds.

\subsection{Case II}
\begin{figure}[htbp]
\centering
\includegraphics[scale=.4]{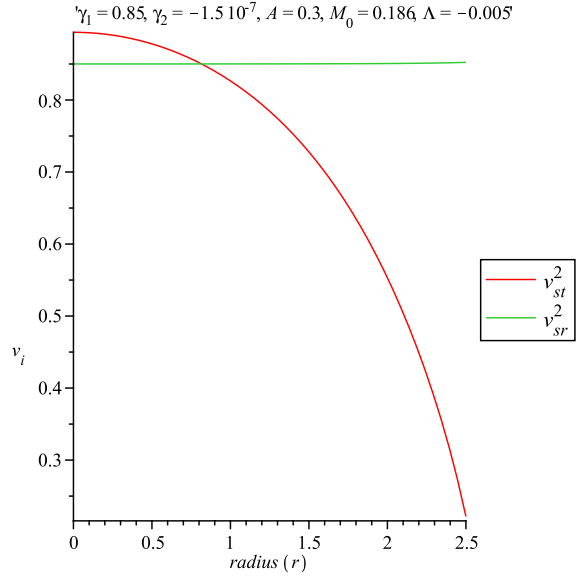}
\caption{Radial and tangential sound speeds plotted against $r$ (Case II).}
\label{fig5}
\end{figure}

From Eqs.~(\ref{eq36})-(\ref{eq39}), we determine
\begin{eqnarray}
v^2_{sr} &=& \frac{dp_r}{d\rho} = \gamma_1-\frac{4\pi^2\gamma_2}{\left(Ae^{-Ar^2}-\Lambda\right)^2},\label{eq43}\\
v^2_{st} &=& \frac{dp_t}{d\rho} = \left[-\gamma_1\left(\left(\gamma_1-1\right)(1-Ar^2)-1\right) +\frac{2\Lambda
(\gamma_1+1)}{A}e^{Ar^2}-\left(3-A(1-2\gamma_1)r^2\right)\left(\frac{4\pi^2\gamma_2}{(Ae^{-Ar^2}-\Lambda)^2}\right)
     \right.\nonumber\\
&& \left. -\frac{e^{Ar^2}}{A}
      \left(\Lambda \gamma_1+\Lambda-\frac{4\pi^2\gamma_2}{Ae^{-Ar^2}-\Lambda}\right)\left((1-2\gamma_1)-\frac{8\pi^2Ar^2}
      {(Ae^{-Ar^2-\Lambda})^2}\right)-\frac{(1+Ar^2)e^{2Ar^2}}{A^2}\left(\Lambda \gamma_1+\Lambda-\frac{4\pi^2\gamma_2}{Ae^{-Ar^2}
      -\Lambda}\right)^2 \right.\nonumber\\
&&
        \left.
         -\frac{2\pi^2\gamma_2e^{Ar^2}}{A^2r}\left(\frac{4Ar}{Ae^{-Ar^2}-\Lambda}+\frac{8A^4r^3e^{-2Ar^2}}{(Ae^{-Ar^2}-\Lambda)^3}\right) \right].\label{eq44}
\end{eqnarray}
Fig.~(\ref{fig5}) shows regular behaviour of radial and transverse sound speeds for the non-linear case as well for specific
choices of the model parameters.

\section{Some features of the model}
\begin{figure}[htbp]
\centering
\includegraphics[scale=.4]{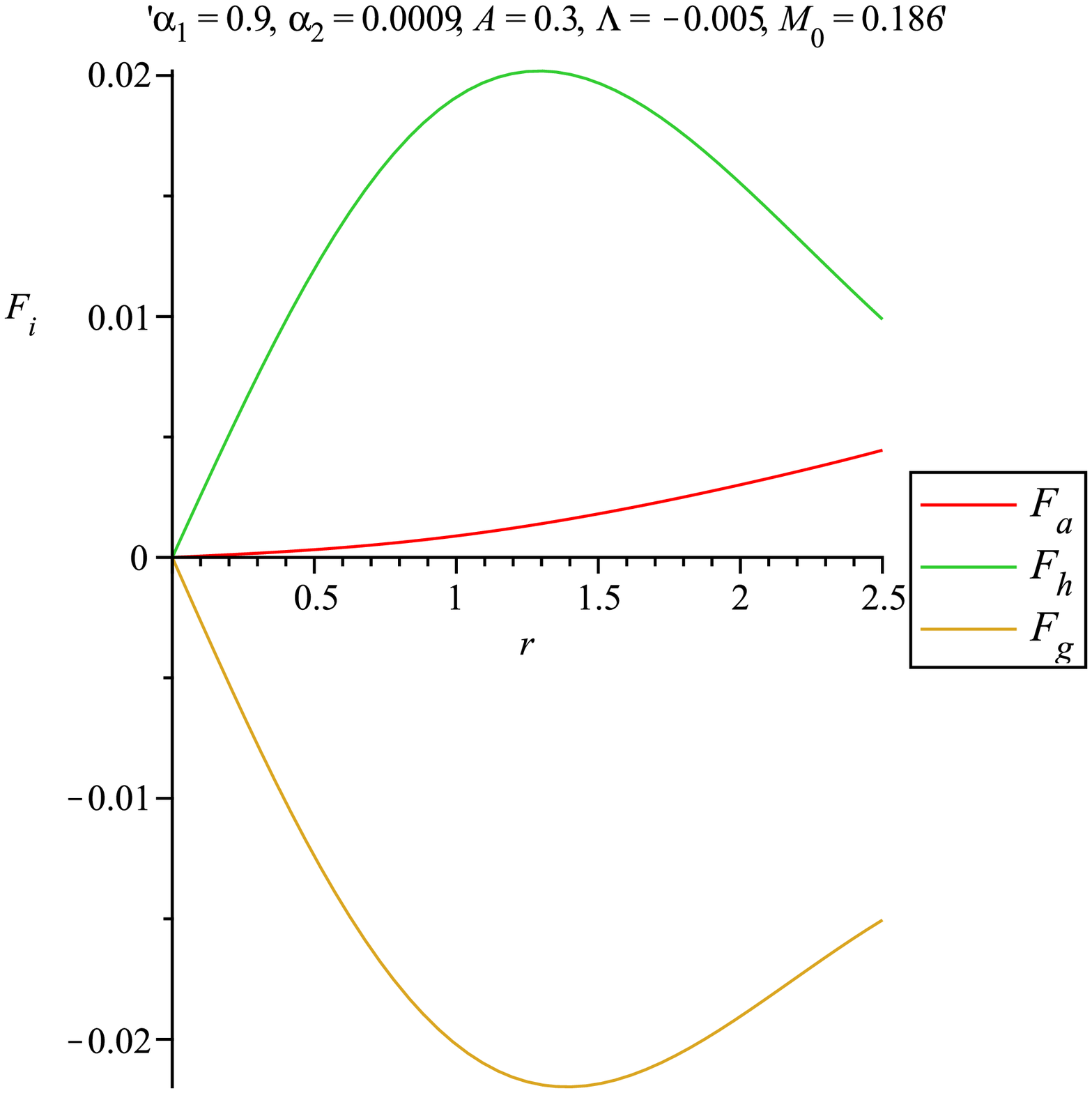}
\caption{Behaviour of three different forces acting on the fluid in static
equilibrium at the stellar interior (Case I).}
\label{fig6}
\end{figure}

\begin{figure}[htbp]
\centering
\includegraphics[scale=.4]{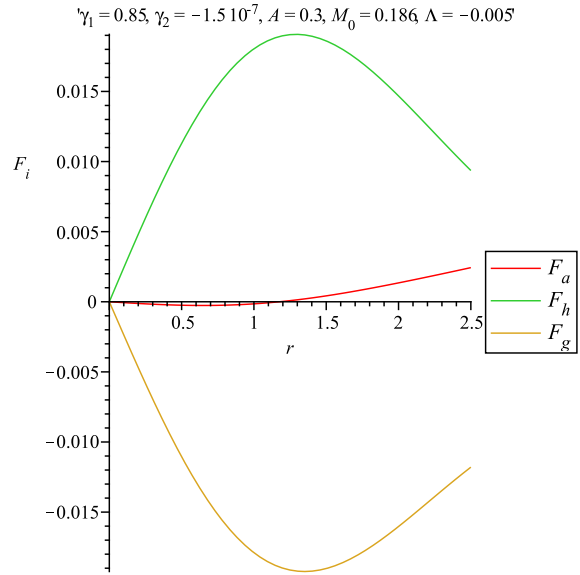}
\caption{Behaviour of three different forces acting on the fluid in static
equilibrium at the stellar interior (Case II).}
\label{fig7}
\end{figure}

We rewrite Eq.~(\ref{eq5}) in the form
\begin{equation}
-\frac{M_G\left(\rho+p_r\right)}{r}e^{\frac{\mu-\nu}{2}}-\frac{d}{dr}\left(p_r
-\frac{\Lambda}{2\pi} \right) +\frac{1}{r}\left(p_t-p_r\right) =
0,\label{eq45}
\end{equation}
where $M_G(r)$ is the  Tolman-Whittaker mass
\cite{Leon1993} and is given by
\begin{equation}
M_G(r) = re^{\frac{\nu-\mu}{2}}\nu^{\prime}.\label{eq46}
\end{equation}
Eq.~(\ref{eq45}) provides the equilibrium condition which implies that the stellar configuration will be in equilibrium under
 the combined impacts of gravitational force ($F_g$), hydrostatic force ($F_h$) and another force term due to pressure anisotropy
 ($F_a$) given respectively by
\begin{eqnarray}
F_g &=& -\frac{M_G\left(\rho+p_r\right)}{r}e^{\frac{\mu-\nu}{2}}, \label{eq47}\\
F_h &=& -\frac{d}{dr}\left(p_r -\frac{\Lambda}{2\pi} \right),
 \label{eq48}\\
F_a &=& \frac{1}{r}\left(p_t -p_r\right). \label{eq49}
\end{eqnarray}
In Fig.~(\ref{fig6}) and (\ref{fig7}), the nature of the three
forces have been shown for Case I and Case II, respectively.

The total gravitational mass $M(r=R)$ in our model can be obtained from Eq.~(\ref{eq7}). Plugging in $C=1$ and $2\mu(r)= Ar^2$ in Eq.~(\ref{eq7}), we obtain
\begin{equation}
m(r) = \frac{1}{2}-\frac{e^{-Ar^2}}{2}-\frac{\Lambda r^2}{2},\label{eq50}
\end{equation}
which can be utilized to determine the compactness of the  star as
\begin{equation}
\frac{M}{R}
 = \frac{1-\Lambda R^2-e^{-AR^2} }{2R}.\label{eq51}
\end{equation}
For $R=2.5$, the compactness $\frac{M}{R}$ is found to be $0.172$  for $\Lambda = -0.002$ and $0.194 $ for $\Lambda = -0.02$.

 The corresponding surface red-shift
\begin{equation}
Z_s = \left(1-\frac{2M}{R}\right)^{-\frac{1}{2}} - 1,\label{eq52}
\end{equation}
turns out be $0.234 $ and $0.279$, respectively.

\section{Discussions}

In this work, we have generated new analytic solutions for a circularly symmetric star which admits a linear or non-linear equation
of state. The matter composition of the star has been assumed to be anisotropic in nature. The values of the constants in our solution
have been fixed by matching the interior solution to the BTZ exterior metric. The cosmological constant $\Lambda$ remains a free parameter
 in our construction. In the absence of any definite value of $\Lambda$, we have made some specific choices of the cosmological constant and shown that the solutions provided here are well behaved and can be utilized to develop physically acceptable model of a static circularly symmetric star in AdS space-time.

\section{Acknowledgments}
FR, KJ and RS would like to thank the Inter-University Centre for
Astronomy and Astrophysics (IUCAA), Pune, India, for awarding Visiting Research Associateship. FR is grateful to UGC, Govt. of India, for financial support
under its Research Award Scheme.

\end{document}